# Controlled Delivery of Reactive Oxygen Species by Functionalized Catalytic Plasma Coatings for Antimicrobial Applications

Paula Navascués[1]*, Flaela Kalemi[1], Flavia Zuber[2], Philipp Meier[3], Ludovica M. Epasto[4], Michal Góra[1,5], Barbara Hanselmann[1], Svetlana Kucher[4], Enrica Bordignon[4], Qun Ren[2], Giacomo Reina[3]*, Dirk Hegemann[1]*.

[1] *Laboratory for Advanced Fibers, Empa, Swiss Federal Laboratories for Materials Science and Technology, St. Gallen, Switzerland.*

[2] *Laboratory for Biointerfaces, Empa, Swiss Federal Laboratories for Materials Science and Technology, St. Gallen, Switzerland.*

[3] *Particles-Biology Interactions Laboratory, Empa, Swiss Federal Laboratories for Materials Science and Technology, St. Gallen, Switzerland.*

[4] *Department of Physical Chemistry, University of Geneva, Geneva, Switzerland.*

[5] *Department of Materials, ETH Zürich, Zürich, Switzerland.*

Corresponding authors (*): paula.denavascues@empa.ch (P.N), giacomo.reina@empa.ch (G.R.), dirk.hegemann@empa.ch (D.H).

**Abstract**

Increasing complications due to bacterial and viral infections require novel antimicrobial approaches. One emerging strategy is that based on catalysts able to selectively deliver reactive oxygen species (ROS) without leaching of other substances. In particular, metal oxide thin films activated by daylight can produce ROS by simply catalyzing oxygen and water molecules. This study examines plasma technology, combining deposition and oxidation processes, as well as plasma polymerization, to obtain functionalized AgOx-doped titanium oxide (TiOx) catalytic materials. The high-energy conditions in the reactive, ionized gas enable the intrinsic formation of a large number of reactive sites at defects and interfaces between the metal oxide nanostructures. Furthermore, plasma functionalization with nanoporous SiOx films (up to 100 nm thick) allows to precisely control the ROS delivery as well as unravel ROS formation mechanism at the metal oxide interface. Combining fluorescence spectroscopy and electron paramagnetic resonance, the controlled delivery of superoxide anion ($O_2^{\bullet-}$) and singlet oxygen ($^1O_2$) has been tuned based on the thickness of the nanoporous functional layer. ROS delivery by functionalized catalytic plasma coating has been related to excellent antimicrobial activity against *E. coli* bacteria as well as murine hepatitis virus, while avoiding cytotoxic and sensitization effects.

**Introduction**

Biocompatible nanocatalysts have emerged over the last decade as a key tool for addressing a variety of global problems. In this context, oxidases mimicking nanostructures have been studied for their catalytic capacity to convert $O_2$ into reactive oxygen species (ROS).[1] ROS include not only free radicals, such as singlet oxygen ($^1O_2$), hydroxyl radical (•OH), and superoxide radical ($O_2^{•-}$) but also non-radical species such as hydrogen peroxide ($H_2O_2$).[2] These highly reactive, oxidizing species enable a range of applications from antimicrobial [3] and cancer treatment [4] to the degradation of organic contaminants such as pollutant removal [5] or water decontamination.[6] The simplicity of ROS compared to complex chemicals such as drugs and other oxidizing substances makes them specific, clean, and green candidates for numerous applications, as they avoid, for example, chlorine or other dangerous side products. In terms of antibacterial and antiviral application, ROS-generating nanomaterials serve as broad-range disinfectants, genuinely beyond antibiotics, making them also ideal candidates for combating antimicrobial resistance (AMR).[3] In this regard, their activity is associated with oxidative stress, damaging cellular components such as lipids, proteins, and DNA, leading to cell death or, at least, growth inhibition.[7] There are plenty of materials capable of producing ROS, mainly due to the catalytic properties of nanomaterials of different dimensions[8] but also strategies such as hydrogels [9] and living-materials.[10] Common strategies in biomedicine, focusing on therapy, are based on ROS production by nanoparticles (i.e., generally 0D materials).[11,12] In this context, ROS are mainly generated by released metal ions (e.g. $Ag^+$) that interfe with the bacterial metabolism. There are, however, fewer studies reporting ROS-producing thin film surfaces for therapeutic approaches, despite such systems have been deeply studied by the photocatalysis community in terms of energy and environmental applications.[13–15] Thus, exploring new approaches by connecting nano-catalysts, biology, and medicine is becoming more and more attractive.[16]

Considering the promising role of ROS for targeted applications, advanced materials capable of producing and precisely controlling the delivery of these active species are desired. ROS can be produced when catalytic active surfaces come into contact with oxygen, and even water molecules.[8] ROS are produced at the catalyst surface and immediately delivered to the environment, rapidly reacting with any substance that can be oxidized. Specific control of the type of ROS produced can be tailored, for example, by applying stimuli such as light [17] or heat.[18] During the last few years, remarkable progress has been made towards the specific production of ROS such as singlet oxygen ($^1O_2$) via light activation of graphene oxide [19] or •OH radicals via Fenton-like reactions (chemodynamic therapy)[16], among other. The controlled diffusion of these radicals into the target media once they are produced, however, has been poorly studied.[20]

Functionalizing ROS-producing surfaces offers a promising strategy for controlling ROS delivery. In this regard, state-of-the-art results generally consist of the incorporation of functional coatings onto $TiO_2$-based nanoparticles via wet chemical methods. For example, K.J. Heo *et al.* reported the functionalization of doped $TiO_2$ nanoparticles with hydrophobic

PFOTES via an aerosol technique to ensure the stability of the system against moisture.[21] Polydimethylsiloxane (PDMS) is the most commonly reported functional polymer used to ensure the mechanical stability and durability of nanoparticles.[5,22] Additionally, R. Ghosh *et al.* recently reported the combination of hydrophilic and hydrophobic functionalization of $TiO_2$ nanoparticles to demonstrate wettability control for water harvesting.[6] Regarding the functionalization of thin film structures, Liu *et al.* published the incorporation of PDMS grafting onto $TiO_2$ thin films.[23] The authors varied the grafting thickness between 0.6 and 5.5 nm and activated the material with UV light. They found a decreasing tendency for the degradation of a generic dye with respect to the grafting thickness, claiming that at 5.5 nm the photocatalytic activity was blocked by the PDMS layer. All these studies demonstrate that catalytic materials can retain at least a degree of activity when subjected to surface functionalization. However, it is not clearly discussed how the activity is tailored and what the mechanisms behind it are. Most of the functional coatings are intrinsically hydrophobic, thereby limiting or delaying the activity of the materials when $H_2O$ is necessary for ROS production. Additionally, due to the high phototoxicity and low tissue permeation depth, UV light should be avoided for clinical applications, justifying the increasing interest in nanomaterials producing ROS by visible and IR light excitation.[18,19] Hence, the preparation of a catalytic, non-metal-releasing thin film with a controlled wide antimicrobial activity is highly desirable for many clinical applications including medical textiles and implants.

Defective $TiO_2$-type thin films are established as a robust solution for photocatalytic applications.[24] Based on that, we reported AgOx-doped TiOx thin films (i.e., AgOx/TiOx) fabricated by low-pressure plasma technology as a rapid antibacterial solution based on ROS generation.[25] The efficacy of this material was tested against gram-positive and gram-negative bacteria, and its activity upon storage in the dark and after daylight reactivation was followed over two years, showing a robust and reproducible behavior avoiding any side effects for human cells.[25] In antimicrobial applications, maintaining ROS homeostasis is crucial: an excessive amount of radicals can activate inflammatory pathways in healthy tissues, while low catalytic efficiency may hinder the sterilization effectiveness of the film.[7] Therefore, understanding and controlling the ROS generation mechanism in AgOx-doped TiOx thin films is decisive for their application. The present study investigates how plasma polymerization can be used to study and control ROS delivery. By depositing functional layers of superhydrophilic nanoporous SiOx (np-SiOx) plasma polymer films, and by ranging the thickness from a few up to 100 nanometers, the amount and type of ROS delivered by the np-SiOx/AgOx/TiOx catalytic nanostructure can be precisely controlled. An in-depth characterization of the mechanisms behind the production and delivery of ROS is performed by using fluorescence spectroscopy and electron paramagnetic resonance. ROS production and delivery have been associated with the antimicrobial activity of the coatings. To the best of our knowledge, this study is the first to report such a precisely quantitative and qualitative controlled release of reactive oxygen species produced at the surface of catalytic thin films activated with daylight.

**Results and Discussion**

Fabrication and functionalization of catalytic plasma coatings (AgOx/TiOx & np-SiOx/AgOx/TiOx)

A previous study by our research group thoroughly investigated on the role of catalytic plasma coatings consisting of AgOx-doped amorphous titanium oxide (TiOx), i.e., AgOx/TiOx, as antibacterial agents based on ROS generation.[25] The plasma sputtering process yields a nonstoichiometric oxidation state of TiOx, stabilized by oxygen vacancies ($O_V$), while the plasma post-oxidation step affixes the AgOx structures, avoiding the release of Ag ions.[26] The AgOx-doping incorporates silver oxide nanoislets, which act as catalytic active sites and narrow the band gap of the material.[27] The activity of the catalytic plasma coating was demonstrated after daylight activation, without the need for specific UV activation or high-intensity lamps. A clear correlation between the production of reactive oxygen species (ROS), as produced when the surface is in contact with water and oxygen, and the antibacterial properties was found, by studying the activity of the material after months of storage in the dark. Building on these previous findings, the current study selects an optimized AgOx/TiOx system as the ROS-producing platform.

**Scheme 1** illustrates the multistep plasma process applied to fabricate the catalytic metal oxides (AgOx/TiOx) and its functionalization with nanoporous plasma polymer films (np-SiOx/AgOx/TiOx). The samples consist of nanoscale thin films produced entirely using low-pressure plasma technology close to room temperature (RT). Specifically, it consists of a 50 nm-thick TiOx layer with nominally 5 nm of Ag deposited on top, subsequently sputtering from titanium and silver targets, respectively (**Scheme 1**, steps 1 and 2). A variety of materials are utilized as substrates, including glass cover slides, silicon wafers, and various textiles/nonwovens such as used for wound dressings. The operation at RT enables the coating of sensitive substrates without inducing any damage, as can be observed in the photograph of the wound dressing on the right side of **Scheme 1** (the gray color corresponds to the AgOx/TiOx coating deposited on the textile). This fact is particularly relevant, taking into account the established application of ROS delivery for wound healing.[7] After plasma post-oxidation (step 3), the AgOx/TiOx catalyst is provided with an average thickness of 55 ± 5 nm. The material has been further functionalized with nanoporous plasma polymer films (PPFs, here np-SiOx) of different thickness to control ROS delivery (step 4). This functionalization was carried out by plasma polymerization and etching using hexamethyldisiloxane (HMDSO) and oxygen.[28,29] Thus, a nanoporous plasma polymer film with SiOx chemistry [29] has been deposited as functional layer varying the thickness between 7 and 100 nm. The PPF shows a superhydrophilic wetting behavior that, together with the interconnectivity of the nanopores, allows the diffusion of water and oxygen molecules through the np-SiOx layer,[29] reaching the plasma polymer-metal oxide interface where ROS are generated. Afterwards, the produced ROS can diffuse through the PPF to the outside, with the functional layer acting as a ROS-delivery control agent. The plasma polymer structure is resistant to highly oxidizing conditions, as those presented in the ROS environment, because of the stability of the siloxane polymer matrix. More details about the fabrication procedure be found in **Methods**.

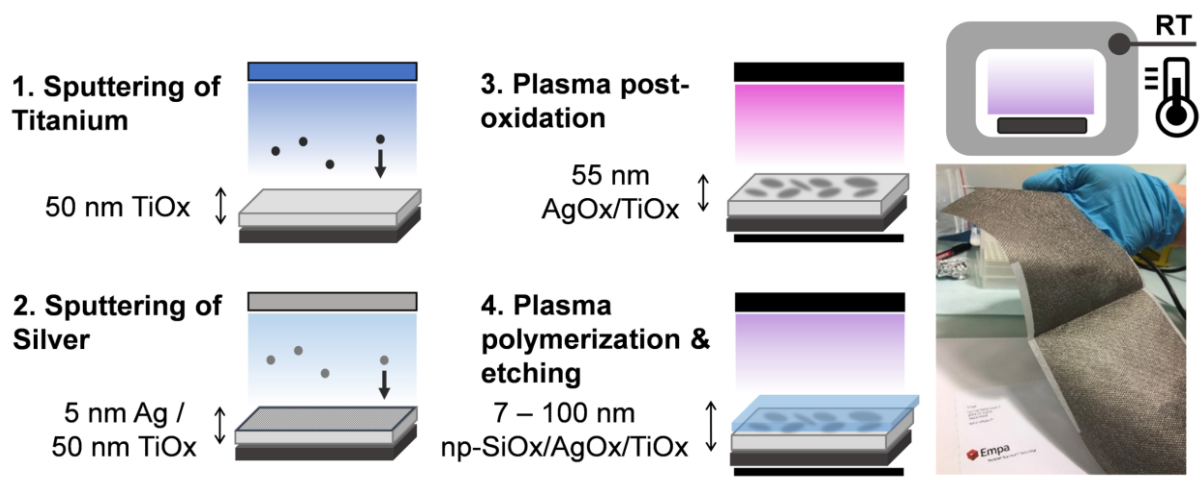

**Scheme 1.** Fabrication of catalytic coatings and its functionalization at RT by a multistep plasma process (steps 1 to 4): (1) sputtering from a titanium target to fabricate the TiOx support, 50 nm-thick, (2) sputtering from a silver target to deposit 5 nm of Ag onto the TiOx support, (3) Ar/O$_2$ plasma post-oxidation to obtain the AgOx/TiOx system, (4) plasma polymerization and etching of HMDSO to functionalize AgOx/TiOx. On the right side, a photograph of a wound dressing coated with the AgOx/TiOx material is also presented.

**Figure 1** presents the catalytic plasma coating's characterization of, including AgOx/TiOx, with functionalization (np-SiOx/AgOx/TiOx), and just the np-SiOx functional layer. Silver oxide nanoislets of different sizes (10 to 100 nm) can be observed dispersed on the TiOx support (see **Figure 1 a.1)**). As mentioned above, these nanointerfaces contribute to the ROS production, acting as catalytic active sites alongside the intrinsic defects.[25] Comparable characterization of the TiOx support alone and the system before plasma post-oxidation (Ag/TiOx) can be found in our previous study.[25] EDS characterization in **Figure 1 b.1)** clearly reveals the homogeneous and flat TiOx support (Ti), the distribution of the silver oxide nanoislets on it (Ag), the superposition of both elements (Ti+Ag), and the overall presence of oxygen at the surface (O). Atomic force microscope (AFM) analysis in **Figure 1 c.1)** allows to distinguish between the roughness of the TiOx support and the AgOx islets on top, revealing an average roughness of 3.4 nm. **Figure 1 a.2)** presents scanning electron microscopy (SEM) (left) and AFM (right) analysis of the np-SiOx/AgOx/TiOx system. The results show that the addition of a thin layer of np-SiOx largely preserves the morphologies previously characterized for the AgOx/TiOx material. The deposition of np-SiOx slightly affects the Ag nanostructures, inducing additional oxidation, with an average roughness of 9.9 nm as determined by AFM. The micrographs shown in **Figure 1 a.2)** correspond to a np-SiOx thickness of 15 nm, but similar features were obtained for coatings examined in the 7-100 nm range. The functional layer is conformal, with no cracks or defects, ensuring H$_2$O and O$_2$ diffusion as well as ROS delivery, is determined by the SiOx intrinsic nanoporosity and its thickness. Additional morphological characterization such as cross section analysis of np-SiOx/AgOx/TiOx can be found in **S2**. **Figure 1 b.2)** shows chemical and (c.2) morphological characterizations of the np-SiOx plasma polymer. ATR-FTIR analysis of the thin film in **Figure 1 b.2)** reveals a dominant SiOx chemistry combined with OH groups due to the superhydrophilic behavior of the surface and the Si-OH pore wall functionalization.[29] The SiO$_2$ band is shifted to the longitudinal optical (LO) vibrational mode at 1228 cm$^{-1}$, related to the nanoporous nature of the material.[30] AFM analysis of the np-SiOx film in **Figure 1 c.2)** indicates conformal coating by a low average roughness of 0.22 nm. Note that the intrinsic nanoporosity of the plasma polymer results from the oxygen etching of sacrificial hydrocarbons in the silica matrix, yielding a pore size ≤ 1 nm [29] and a very low surface roughness.

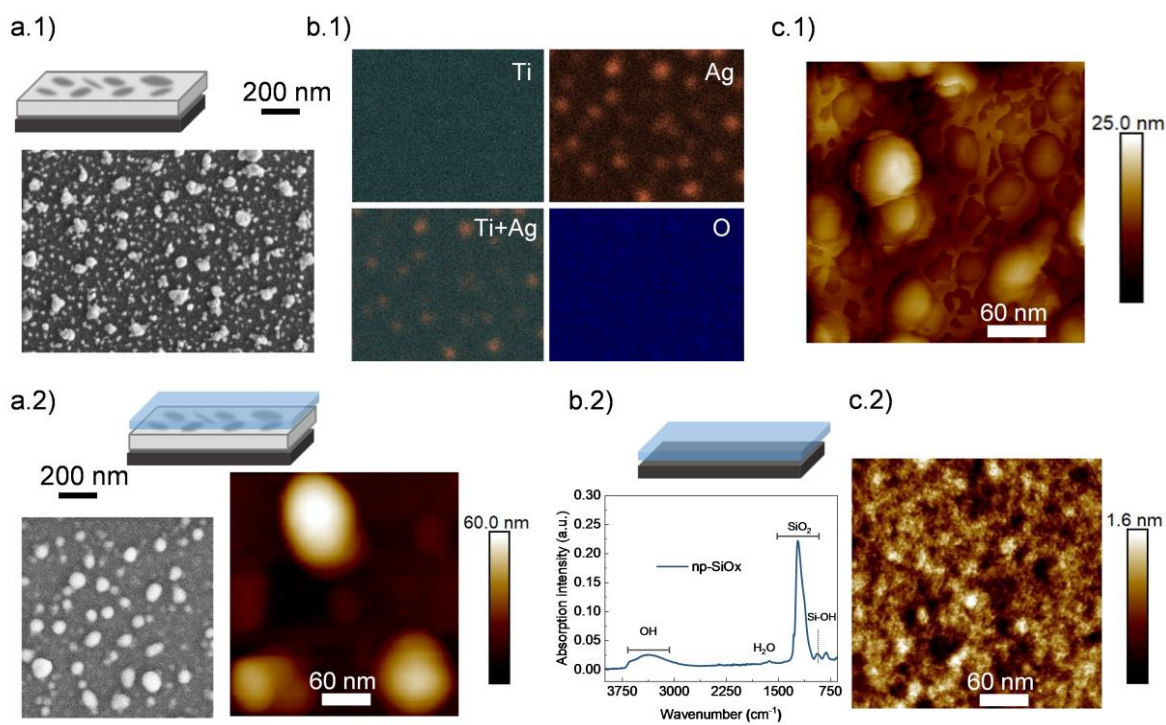

**Figure 1.** Characterization of the ROS-producing system (i.e., the catalytic AgOx/TiOx plasma coating) and its functionalization with nanoporous plasma polymer films. a.1) SEM top view micrograph of AgOx/TiOx. b.1) EDS analysis of AgOx/TiOx, including composition maps of titanium (Ti), silver (Ag), Ti and Ag overlapped (Ti+Ag), and oxygen (O). c.1) AFM micrograph of AgOx/TiOx. a.2) np-SiOx/AgOx/TiOx surface analyzed in top view by (left) SEM and (right) AFM. b.2) ATR-FTIR characterization of np-SiOx thin film and c.2) AFM high resolution analysis of its smooth surface.

Absolute UV-Vis optical characterization (300-900 nm, **Figure 2**) of the catalytic plasma coating and its functionalization was recorded with an integrating sphere. **Figure 2 a)** shows the absorptance (i.e., the percentage of light not transmitted nor reflected) of the different layers concerning the fabrication process of AgOx/TiOx. Compared to standard anatase TiO$_2$ materials,[31] the nonstoichiometric TiOx layer is already characterized by light absorption in the full UV-Vis range and not only in the UV. This is due to the oxygen vacancies introduced during plasma processing,[26] retarding also charge recombination.[32] In this regard, **Figure 2 a)** shows effects of tuning the band gap by adding Ag doping and plasma post-oxidation. Additionally, transmittance measurements of the np-SiOx layer indicate its transparency with transmittance of approximately 90% and no specific absorption bands, overlapping with the transmittance of a glass substrate (see **Figure 2 b)**). This result points out that there is no light loss due to the addition of the functional np-SiOx layer. Indeed, an opposite effect has been detected: **Figure 2 c)** gives absorptance and total reflectance curves for an AgOx/TiOx thin film and the same film covered by a 7 nm functional layer (np-SiOx/AgOx/TiOx). All curves follow the same behavior, indicating no changes in the structure of the band gap, but with an offset positive contribution for the absorption of the functionalized sample with respect to the non-coated catalyst. Certainly, the np-SiOx thin layer acts as an anti-reflective coating,[33] as indicated by the opposite tendency observed in total reflectance measurements (see dashed plots in **Figure 2 c)**). No differences in total transmittance or the diffuse components were detected, as shown in **S2**. Indeed, this antireflective behavior is expected considering the low refractive index of the np-SiOx film (n=1.43, according to ellipsometry measurements). A similar optical behavior is expected for thicker functional np-SiOx layers, since the thickness range varied in this study (7-100 nm) is generally below one-quarter of the wavelength of the incident light.[33]

Small differences in absorptance behavior can be observed by comparing the AgOx/TiOx curves in **Figure 2 a)** and **c)**, respectively. These are due to small differences in thickness in the 55 ± 5 nm range. Nearly negligible values of the spectral diffuse components indicate a high optical quality of the plasma coatings (see **S2**). The results reported above demonstrate the transparent and anti-reflective behavior of the np-SiOx functional layers. Additionally, their conformal and low-roughness morphologies, combined with the nanoporous superhydrophilic nature and oxidation-resistant SiOx chemistry, make them ideal candidates for controlling ROS delivery based on nano-scaled polymer thickness.

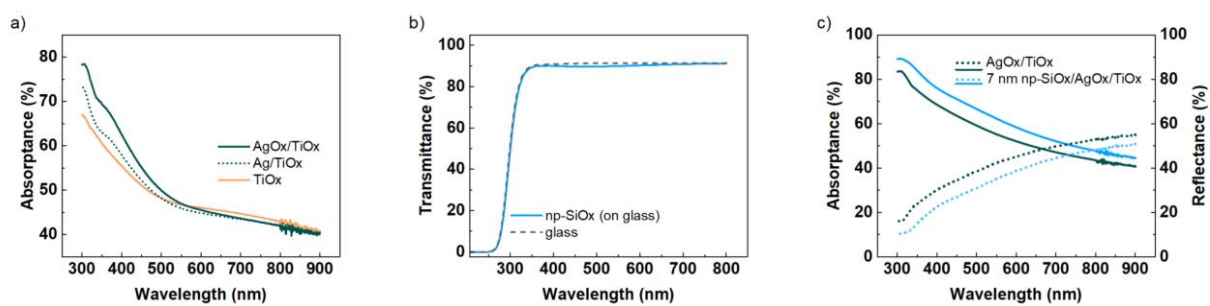

**Figure 2.** UV-Vis characterization of catalytic plasma coatings. a) Absorptance spectra of the different layers of the catalytic metal oxide (TiOx support, Ag/TiOx and AgOx/TiOx; thickness 55 ± 5 nm). b) Similar transmittance of (continuous line) np-SiOx deposited on glass and (dashed line) glass substrate. c) Comparative (continuous line) absorptance and (dashed line) reflectance spectra of AgOx/TiOx and np-SiOx/AgOx/TiOx plasma coatings, showing the anti-reflective behavior of the np-SiOx layer (7 nm np-SiOx, 55 ± 5 nm AgOx/TiOx).

Production and delivery of ROS by functionalized catalytic plasma coatings

In our previous study, the estimation for the ROS production was related to the antibacterial efficacy of the AgOx/TiOx material after exposure to daylight.[25] Indeed, the UV-Vis absorptance characterization (see **Figure 2**) demonstrates that the catalytic plasma coating absorbs light in the visible range, providing the necessary energy to activate the catalyst, while the nanostructures shown in **Figure 1** act as catalytic active sites for the reactions to take place. When $H_2O$ and $O_2$ molecules reach these interfaces, short-living reactive oxygen species are formed (•OH, $O_2^{•-}$), which can recombine afterwards to long-living $H_2O_2$. **Figure 3** presents general and specific detection of different radicals for the AgOx/TiOx catalytic plasma coating and its functionalization with the nanoporous SiOx layer. Overall ROS detection levels are measured by following the ROS-induced oxidation of dihydrorhodamine 123 (DHR 123) into the fluorescent rhodamine 123 (R 123), shown in **Figure 3 a)**. Fluorescence signal is normalized with respect to the control DHR 123 solution. A tendency between the polymer thickness and the rhodamine detection can be observed, suggesting that the release of radicals is tuned by controlling the thickness of the np-SiOx layer. The maximum activity is detected for the sample with the thinnest (7 nm) functionalization, then decreasing with increasing np-SiOx thickness, with no significant rhodamine detection for functional layers thicker than 60 nm. In this latter case, the ROS signal is similar to that of the glass (substrate) control. The higher activity of the 7 nm SiOx functionalized sample with respect to the bare catalyst (i.e., bare AgOx/TiOx) can be related to different factors: i) the superhydrophilic (contact angle below 10°) wetting behaviour of the np-SiOx layer compared to the less hydrophilic AgOx/TiOx surface (approximately 60°), ii) a possible additional activating oxidation effect of the catalyst, during plasma functionalization, slightly affecting its morphology (see **Figure 1 a.2)**), and iii) the higher light absorption shown in **Figure 2 c)**. However, the latter point should play a minor role, as fluorescence measurements are carried out in the plate reader (i.e., in the dark), only exposing the material to the 488 nm excitation wavelength for R 123 detection. Moreover, the possibility of a further oxidation of R 123 cannot be discarded, which would avoid its detection, decreasing overall ROS levels. This tendency found between ROS delivery and the polymer thickness can be ascribed to the lifetime and the recombination probability of reactive species within the nanoporous SiOx channels. Longer mean free paths through the polymer matrix may promote ROS recombination at the pores, even for long-living $H_2O_2$, reducing their overall oxidative activity. Besides, DHR 123 is a common probe for detecting reactive species, providing a general idea of the oxidative capability of a system, but not specific information about the involved radicals.

In order to shed light on the mechanisms, an additional experiment was performed with dihydroethidium (DHE), a specific probe for $O_2^{•-}$ (i.e., superoxide) detection. Superoxide is an important radical ion formed from the single electron reduction of $O_2$ molecules and effective against different kinds of bacteria. Kinetic measurements of DHE degradation reveal a clear correlation between the oxidative capability of the system and the SiOx thickness (**Figure 3 b)**). In particular, by fitting the curves based on a pseudo-first order reaction kinetic, a reproducible decrease in the reaction constant is obtained: for a reaction constant (~$1.4·10^{-3}$ $s^{-1}$) for the AgOx/TiOx plasma coating, it is reduced by 35%, 67%, and 72% for 7, 15, and 30 nm np-SiOx functionalizations, respectively. For np-SiOx functional layers thicker than 60 nm, a very small activity is detected, with a kinetic comparable to that of the glass substrate, similar to what has been shown for DHR 123 (**Figure 3 a)**). Additional details regarding DHE kinetic degradation can be found in **S3**.

To confirm that superoxide was generated by the reduction of $O_2$, the same experiment was performed in anoxic (no $O_2$) conditions (see **S4** for details). In these conditions, the drop in oxidation efficiency of the catalytic plasma coating confirms that the process is $O_2$ mediated. Moreover, the production of superoxide radicals was confirmed with an inhibition experiment performed with glutathione (GSH, see **S5**). GSH is a major non-enzymatic ROS scavenger preferentially oxidized in presence of $O_2^{\bullet-}$; this antioxidant property plays a key role in protecting mammalian cells from ROS-induced oxidative damage.[34] Therefore, when added to a DHE solution, GSH can compete with the dye to be oxidized by $O_2^{\bullet-}$. A clear inhibition of DHE degradation is observed for GSH concentrations more than ten times higher than that of DHE, confirming that that $O_2^{\bullet-}$ is generated and can be scavenged by GSH. Overall, our results suggest that superoxide is formed from reduction of molecular oxygen and that their delivery can be controlled by the thickness of the np-SiOx layer. In terms of superoxide production, the higher concentrations are detected for the AgOx/TiOx and 7 nm np-SiOx/AgOx/TiOx systems, able to fully oxidize DHE in one hour. Taking into account that the concentration of the fluorescent probe was 10 µM in 300 µL (i.e., 3 nmol), that the diameter of the samples was 1 cm (i.e., surface of 0.785 cm$^2$), and the stochiometry between $O_2^{\bullet-}$ and DHE is 1:1, we can conclude that the production of superoxide by active plasma coatings is at least 0.38 µmol·h$^{-1}$·cm$^{-2}$ in normoxic conditions. Based on the GSH inhibition experiment, this production rate may be more than ten times higher. We can thus confirm that a strong superoxide formation efficiency is present, with kinetics similar to other photocatalytic systems [35] but without the need for light stimuli during the experiment. As mentioned above, the catalyst is active after light exposure, but the analyzed fluorescence experiments were performed in dark conditions.

To further investigate the type of ROS produced, another specific fluorescent probe has been used. **Figure 3 c)** presents an experiment carried out with a 9,10-Anthracenediyl-bis(methylene)dimalonic acid (ABDA) solution. ABDA is a probe that is specifically oxydized by singlet oxygen ($^1O_2$) into a non-fluorescent product.[36] Therefore, singlet oxygen production was monitored by following the decrease in fluorescence intensity.[37] The AgOx/TiOx sample does not result in a loss in fluorescence, with similar behavior as the glass substrate, revealing there is no $^1O_2$ production by the catalytic plasma coatings. Surprisingly, a clear decrease in the fluorescence signal is detected for the sample functionalized with the thinnest functional polymer (i.e., 7 nm np-SiOx/AgOx/TiOx), but no decrease for thicker functional layers (i.e., $d_{np-SiOx}$ > 7 nm). Different mechanisms can be discussed for $^1O_2$ production with TiO$_2$-based catalytic materials, mainly the direct production by energy transfer between $O_2$ and a triplet state of the material [38] and via the oxidation of $O_2^{\bullet-}$ at the holes.[39] We propose that the addition of the nanoporous functional layer increases the residence time of superoxide radicals at the catalyst, promoting its oxidation to $^1O_2$ at the holes. Taking into account the absence of $^1O_2$ detected for the AgOx/TiOx sample, the $^1O_2$ direct production from $O_2$ may not play a role. The oxidation of $O_2^{\bullet-}$ should similarly happen for thicker plasma polymer functionalization. However, singlet oxygen is characeized by a shorter lifetime in water (microseconds at 25°C[38]), compared to other ROS under similar conditions such as $O_2^{\bullet-}$ (miliseconds[40]) or $H_2O_2$ (hours to days[41]). Therefore, no ABDA degradation is detected for $d_{np-SiOx}$ > 7 nm, i.e., there is no delivery of $^1O_2$ for thicker functionalizations. These results agree with the highest overall ROS detection by DHR in **Figure 3 a)**. Therefore, the introduction of a plasma polymer functionalization allows precise tunning of the type of ROS produced. Importantly, $^1O_2$ can be detected in the overall ROS picture by applying a nano-scaled functionalization, which is a desired reactive species for many applications due to its strong oxidative properties.[38]

It is important to remark that the used fluorescent probes, i.e., DHR 123, DHE, and ABDA, are considerably large molecules with several aromatic rings. They are too large (>1 nm) to penetrate and diffuse with $H_2O$ through the nanoporous network (≤1 nm). Hence, they are not directly oxidized by the active surface (i.e., the AgOx/TiOx surface). ROS measured by fluorescence spectroscopy are detected outside the nanoporous plasma polymer, demonstrating the ROS-delivery capability of plasma functionalization. A different situation occurs for the following experiment aiming to detect •OH radicals. **Figure 3 d)** shows continuous wave electron paramagnetic resonance (cw EPR) spectra with 5,5-Dimethyl-1-pyrroline *N*-oxide (DMPO) present in solution as a spin trap for ROS. The presence of hydroxyl radicals is detected for both AgOx/TiOx and np-SiOx/AgOx/TiOx systems. After 44 minutes of incubation, the highest signal is obtained for the 7 nm np-SiOx/AgOx/TiOx sample, while similar overall intensities are obtained for the bare catalyst and thicker functionalizations (see **S6** for complementary data). It is important to remark that, in this case, to accelerate the ROS production reaction kinetic, experiments were performed while incubating the sample under light exposure (common LED lamp). This might explain why the 7 nm np-SiOx/AgOx/TiOx sample has a relatively stronger signal, due to the antireflective properties of the coating, discussed in **Figure 2 b)**. To confirm that the DMPO$^·$-OH signals arise from direct trapping of •OH by DMPO, we added DMSO to the aqueous solution, which is a known scavenger of •OH radicals. After prolonged incubation in presence of DMSO, we found a negligible DMPO$^·$-OH signal, proving that the EPR signals detected in aqueous buffers (**Fig. 3 d)**) specifically probe the formation of •OH radicals and are not caused by degradation of short-lived DMPO$^·$-OOH adducts into DMPO$^·$-OH (see **S7**). In contrast to DHR 123, DHE, and ABDA, the DMPO molecule is much smaller in size, enabling it to traverse through the porous network and reach the AgOx/TiOx surface where catalytic reactions are taking place. For this reason, •OH radicals are detected for all the tested functionalized samples (7, 15, and 30 nm np-SiOx). Otherwise, they should not have been detected, taking into account the short lifetime of these radicals (nanoseconds), generally scavenged a few angstroms from its generation site.[42] The enhanced detection of •OH radicals with DMPO by EPR (see **Figure 3 d)**) should trigger a higher concentration of $H_2O_2$ for the 7 nm np-SiOx functionalization compared to the plain catalyst (AgOx/TiOx). This, together with the singlet oxygen detection shown in **Figure 3 c)**, agrees with the higher overall ROS detection by following the oxidation of DHR 123 (see **Figure 3 a)**) for the 7 nm functionalization.

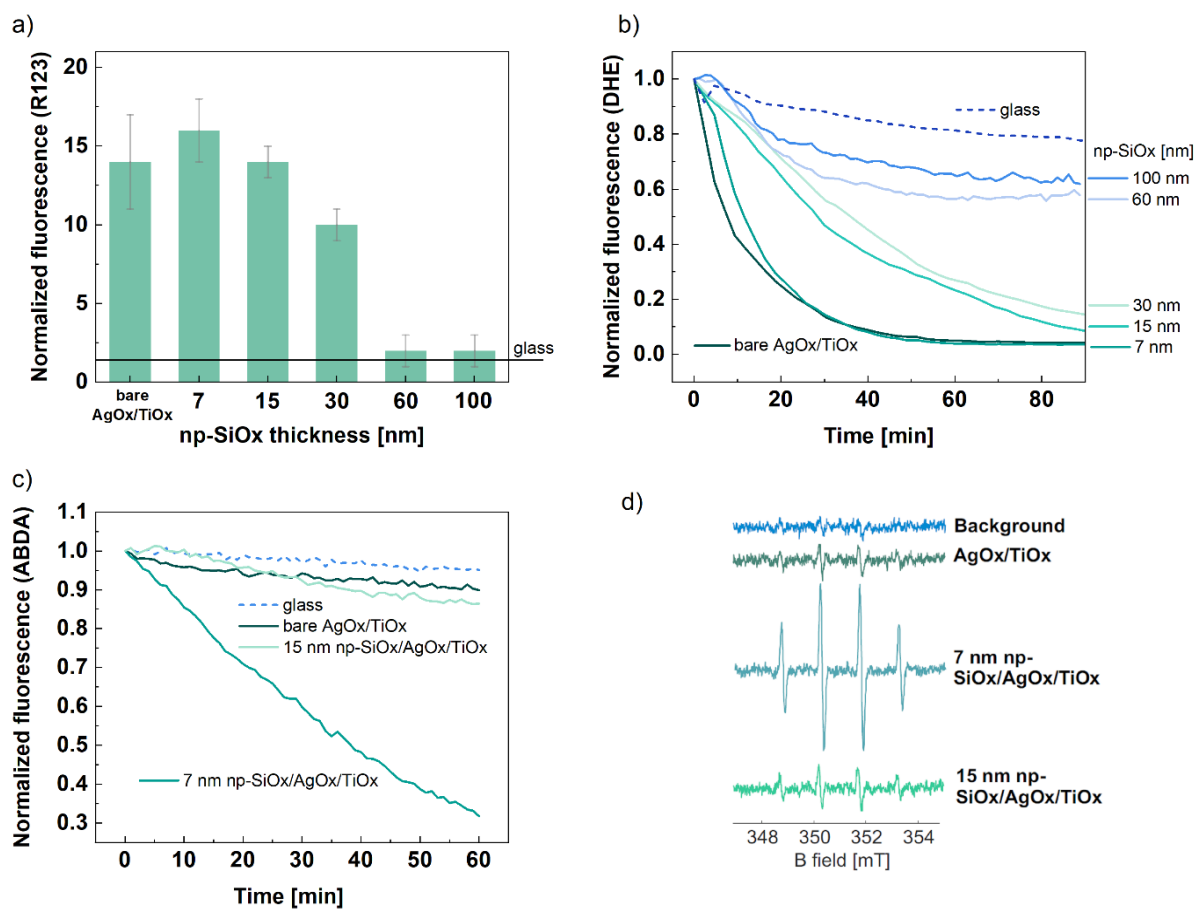

**Figure 3.** Detection of reactive oxygen species (ROS) produced and released by functionalized catalytic plasma coatings by (a)-c)) fluorescence spectroscopy and d) electron paramagnetic resonance (EPR). a) General detection of ROS by following R 123 detection at 35 min of incubation (10µM-DHR 123 solution). b) Specific detection of $O_2^{\bullet-}$ by following DHE degradation (10 µM-DHE solution). c) Specific detection of $^1O_2$ by following the degradation of ABDA (2.4 µM-ABDA solution). d) Specific detection of •OH radicals by EPR using 100 mM DMPO in 20 mM HEPES, 150 mM NaCl buffer at pH 7.3. The spectra shown are collected after 44 minutes of incubation at room temperature. Control experiments were performed on a glass substrate (background) under the same experimental conditions. Additional information can be found in **S6** and **S7**.

The mechanisms of $O_2^{\bullet-}$ and •OH radical generation involves electrons and holes separated in the catalyst, where electrons reduce electron acceptors (molecular oxygen, yielding $O_2^{\bullet-}$), and holes oxidize electron donors including adsorbed water ($H_2O$) and hydroxides ($OH^-$) to give hydroxyl (•OH) radicals.[43,44] Generally, $O_2^{\bullet-}$ and •OH radicals can further react to yield $H_2O_2$.[45] **Scheme 2** illustrates the mechanistic view from **Figure 3**, with ROS detection ($H_2O_2$, $O_2^{\bullet-}$) for the bare AgOx/TiOx catalyst and controlled ROS delivery when including the functional layer (np-SiOx/AgOx/TiOx). Precise control of the np-SiOx thickness allows to specifically deliver $^1O_2$ as well as the controlled release of longer-living reactive species. This capability of controlling ROS delivery represents a promising strategy for therapeutic applications. It goes a step further compared to other approaches such as the use of ROS gels for a slow release over an extended period of time, [3] enabling sustained continuous release of ROS to a target site. Taking into account the nanoporosity of the SiOx layer and its superhydrophilic surface, water and oxygen need only seconds to penetrate and hydrate the layer (see right sketch in **Scheme 2**).[46] Results obtained with the added np-SiOx layer indicate that the accesibility of the catalytic active sites is maintained at the covered interface. Once produced at the interface, ROS diffuse to the outside through the nanoporous network, while they might recombine among themselves and with -OH groups at the pore walls (functionalized with Si-OH groups[29]). For np-SiOx layers thicker than 30 nm, $O_2^{\cdot-}$ is no longer detected, probably indicating that most of ROS have recombined to form $H_2O_2$. Given the small size of DMPO molecules, it is not possible to assess •OH delivery, although these radicals should recombine within the first few nanometers, ultimately forming $H_2O_2$. As mentioned, the thin np-SiOx functionalization of 7 nm allows to specifically deliver $^1O_2$. We performed further measurements to assess the role of the porous volume of the plasma functionalization. Comparative ROS values were detected for np-SiOx layers with a porous volume of 13 ± 2 % and 20 ± 5 %. These negligible differences indicate that the layer thickness is the rate-limiting step rather than the porous volume of the coating.

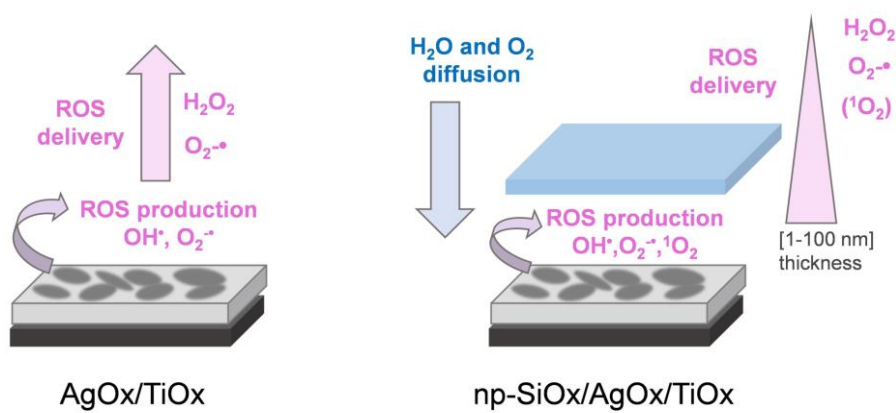

**Scheme 2**. ROS production (AgOx/TiOx) and controlled delivery (np-SiOx/AgOx/TiOx) based on the thickness of the functional nanoporous layer.

Production and delivery of ROS below cytotoxicity and sensitization levels

One key aspect of the catalytic plasma coating for biological and medical applications is the absence of cell cytotoxicity. The AgOx/TiOx material exhibited no cytotoxicity itself, as confirmed by the results in **Figure 4 a)**, using normal human dermal fibroblasts (NHDFs, C-12352, PromoCell) following the ISO10993-5 norm. Additionally, the effect of adding the np-SiOx functional layer was found to induce no cytotoxicity. A positive control consisting of a 1% Triton X-100 solution in DMEM is included in **Figure 4 a)** to properly control cell viability. Furthermore, to test the potential of the catalytic plasma coating for skin treatments, cell viability and sensitization have been tested in artificial sweat. For these tests, the catalytic plasma coatings are directly deposited on wound dressings as shown in the photograph in **Scheme 1**. This test is primarily designed to study the release of ions in an artificial sweat solution, an important factor to consider for nanomaterials composed of metal oxides as in the present case. In accordance with our previous publication,[25] no significant release of ions is detected, as reported in **Figure 4 b)** and **c)**, showing no effect on skin cell viability or sensitization (fold induction) according to OECD guideline 442D, respectively, for both AgOx/TiOx and np-SiOx/AgOx/TiOx samples, whereas positive controls consisting of ethylene glycol dimethacrylate (EGDMA, cell toxic and sensitizer) indicate a high power of sensitizing and irritation. No evident effect can be ascribed to the catalytic plasma coatings, hence, no adverse events for advanced skin cultures are expected. Accordingly, these coatings are applicable in biomedical cases such as treating skin infections or coating external surfaces that may come into contact with living beings (face masks, surgical instruments, or critical surfaces in hospitals, among others) without adverse outcomes for the skin.

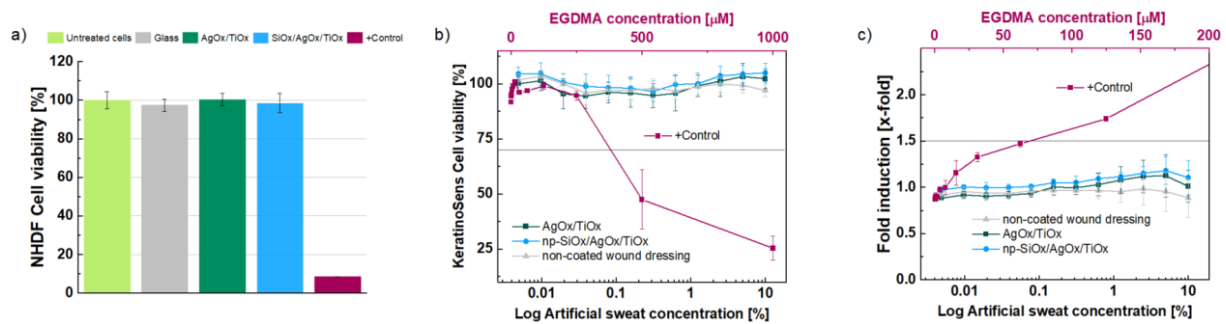

**Figure 4.** a) Cytotoxicity tested with normal human dermal fibroblasts (NHDF) using MTS assay, b) KeratinoSens™ skin irritation (cell viability) and c) sensitization (fold induction) assay. The np-SiOx layer has a 15 nm thickness for both tests. Positive controls (i.e., +Control) are a) 1% Triton X-100 and b)-c) EGDMA. Substrates are also tested, as negative controls, a) glass and b)-c) non-coated wound dressing.

Antibacterial activity of the functionalized catalytic plasma coatings

The catalytic plasma coatings were qualitatively analyzed for their antibacterial activity aginst the gram-negative strain *E. coli* **(Figure 5 a)**) and the gram-positive strain *S. aureus* **(Figure 5 b)**) using a touch test. As can be observed in the figure, the negative control glass substrate did not impact the growth. Against *E. coli*, both the AgOx/TiOx and 7 nm np-SiOx/AgOx/TiOx samples revealed clear growth inhibition already after 10 minutes, with further enhancement observed after 60 minutes. A minor inhibition compared to the glass negative control is detected for the sample functionalized with a thicker (60 nm) np-SiOx coating, slightly increasing after 60 min. A generally higher activity is inferred for the plain catalyst, indicating a slightly delayed activity when the thin plasma polymer is added. The data obtained here are consistent with those discussed above, that is, thicker (>60 nm) np-SiOx layers hindering ROS delivery **(Figure 3)** ultimately resulting in limited bacterial inhibition. Against *S. aureus* **(Figure 5 b)**), the antibacterial activity of the coatings is less pronounced, which aligns with earlier findings [25,47] related to the inherent resistance of Gram-positive bacteria due to their thicker peptidoglycan layer.[47,48] Nevertheless, the same tendency in inhibition as for *E. coli* (c.f. **Figure 5 a)**) is detected: a stronger effect after 60 min compared to 10 min of interaction, particularly for the 7 nm np-SiOx/AgOx/TiOx sample.

To quantitatively analyze the antibacterial activity, a modified ASTM E2180 method was employed to assess the efficacy against the same bacterial strains. **Figure 5 c)** and **d)** present the obtained results for different exposure times (10, 30, and 60 min) for the AgOx/TiOx plasma coating and one functionalization thickness of 15 nm np-SiOx. This intermediate thickness has been selected with the aim of quantitatively detecting a delay in the antibacterial properties of the system. For *E. coli* (**Figure 5 c)**), the functionalized sample seems to be similarly active as the bare catalyst. Note that superhydrophilic surfaces can affect bacteria adhesion because the water layer acts as a barrier,[49] which might counterpart a lower dose of ROS. As can be observed in **Figure 5 c)**, after 10 minutes, both coatings demonstrate similar antibacterial efficacy of 2 logs reduction in colony-forming units (CFU). Extending the interaction time, CFU further reduce, reaching after 60 min over 4 logs of reduction for the functionalized sample and slightly less than 4 logs for the AgOx/TiOx samples. These results indicate that sufficient ROS generation is achieved for effective *E. coli* inhibition in both the AgOx/TiOx and np-SiOx/AgOx/TiOx systems. On the contrary, almost no activity is detected for *S. aureus* by this method, as shown in **Figure 5 d)**, for both bare and functionalized catalytic plasma coatings. It can be concluded that the level of ROS diffusion as diluted in the solution is still sufficient for the inhibition of the used *E. coli* strain but not for the *S. aureus* one.

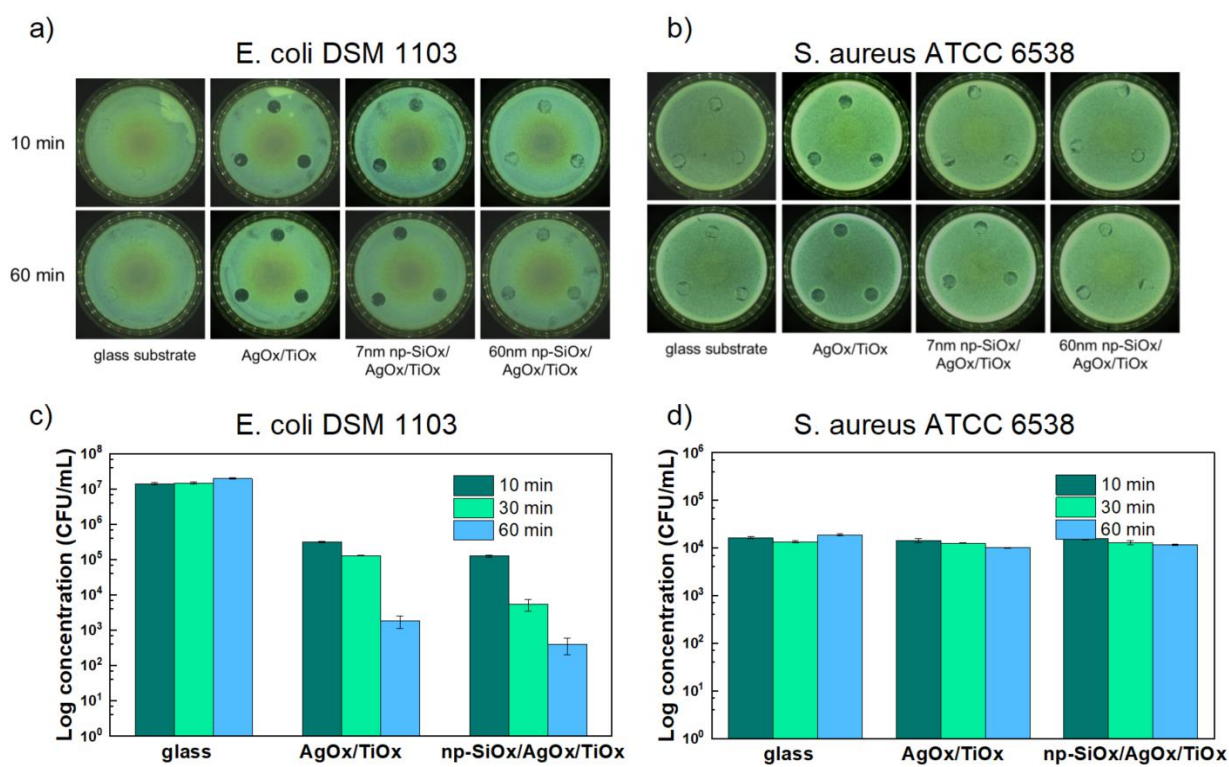

**Figure 5.** Antibacterial capability of the catalytic plasma coatings. a)-b) Qualitative analysis of the antibacterial property of AgOx/TiOx and the functionalization with 7 nm and 60 nm np-SiOx against (a)) *E. coli* DSM 1103 and (b)) *S. aureus* ATCC 6538, for 10 and 60 min of interaction. c)-d) Quantitative analysis of the antibacterial property of AgOx/TiOx and a functionalization with an intermediate thickness of 15 nm for (c)) *E. coli* DSM 1103 and (d)) *S. aureus* ATCC 6538 at three different contact times of 10, 30, and 60 min.

Antiviral activity of functionalized catalytic plasma coatings and underlying ROS mechanism

The antiviral properties of AgOx/TiOx and functionalized np-SiOx/AgOx/TiOx plasma coatings have been investigated by using two different tests regarding the interaction between the produced ROS and two virus models. **Figure 7 a)** gives results for an InVIS assay.[50] This test involves quenching a fluorescent rhodamine-18 probe within the viral membrane of an inactivated Brisbane/2007 influenza virus model. When the virus envelope is intact, no fluorescence is detected, but upon disintegration of the viral capsule, the fluorescence signal of the rhodamine can be measured. AgOx/TiOx and 7 nm np-SiOx/AgOx/TiOx plasma coatings and a glass substrate were tested, the latter included as negative control. For positive controls, the incubation of the virus, together with the samples, is mixed with an octaethylene glycol monododecyl ether OEG detergent that disintegrates the virus by membrane dissolution and protein denaturation.[50]

As can be observed in **Figure 7 a)**, only a low baseline fluorescence signal is detected for both the glass, AgOx/TiOx, and 7 nm np-SiOx/AgOx/TiOx plasma coatings, indicating that the virus membrane remains intact, whereas adding the detergent to all samples induces the charasteristic fluorescence due to viral envelope disintegration. The release of metal ions such as copper ions [51] or the sharp edged structure of graphene oxide [52] have been claimed as possible agents to affect the viral capsid integrity.[53] The absence of ion release and the low, smooth surface roughness of the plasma coatings justify the results in **Figure 7 a)**.

Nevertheless, as for bacteria, there are different possible mechanisms responsible of virus inactivation and destruction aside membrane breakage such as reactions with targeted proteins[54] and ROS diffusion thorugh the membrane [55]. Indeed, 2 and 3 logs of reduction of virus concentration can be seen in **Figure 7 b)** for AgOx/TiOx and 7 nm np-SiOx/AgOx/TiOx samples, respectively, after two hours of incubation of an active murine hepatitis virus according to ISO 18184 with respect to the negative control (i.e., glass substrate after 2 h). It is relevant to study comparable times of incubation, since the viability of a virus on a surface is reduced over time.[56]

Certainly, performing this test for materials deposited on wound dressings, as in **Figure 4 b)** and **c)**, gives no significant virus concentration on both substrates (i.e. dressings) and coated dressings after two hours. This is due to the unstable behavior of viruses on textiles, thus ideal substrate materials for antiviral properties.[57] Therefore, results in **Figure 7 b)** are tested for catalytic plasma coatings deposited on glass, to ensure that the concentration reduction after two hours is due to the material's activity (i.e., ROS release). High antiviral activity is detected for functionalized and non-functionalized samples, higher in the former casedue to the higher general ROS detection for the 7 nm np-SiOx functionalization and the specific delivery of $^1O_2$. Indeed, singlet oxygen has been reported as a major oxidizing agent for viral capsid proteins.[58] Hence, the antiviral efficacy of the material is demonstrated as a result of ROS diffusion through the virus envelope (i.e., lipid membrane) and the reaction with capsid proteins and DNA[54] rather than breaking the virus membrane.

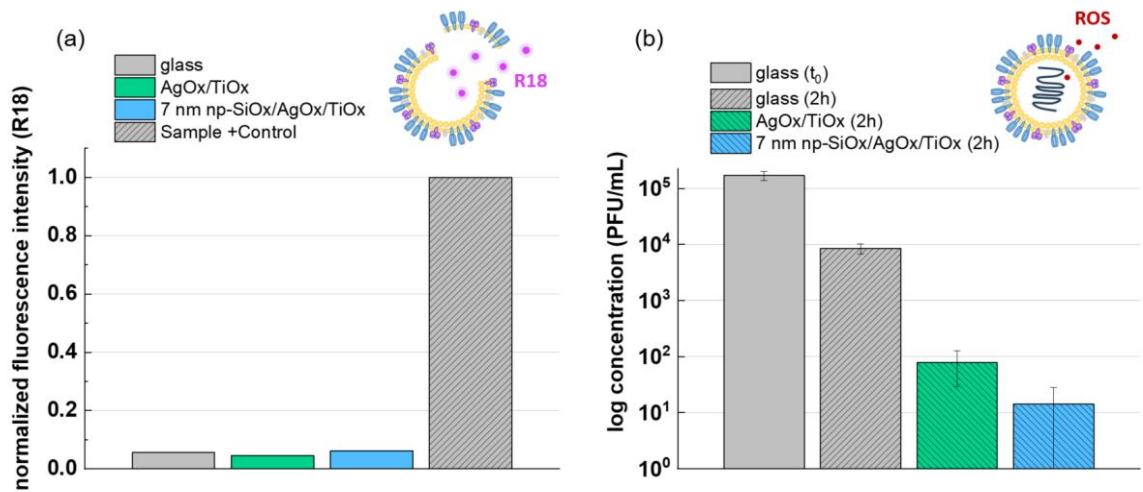

**Figure 7.** Antiviral tests for AgOx/TiOx and 7 nm np-SiOx/AgOx/TiOx plasma coatings. (a) InVIS rhodamine-18-quenched virus experiment, for each sample including a positive control (adding an OEG detergent). (b) Murine hepatitis virus (MHV) inhibition performed according to ISO 18184 plaque assay. Glass substrates were used as negative control.

We can conclude that a relevant production of ROS from catalytic plasma coatings has been identified able to inactivate bacteria and virus based on radical chemistry (i.e., ROS delivery).

**Conclusions**

The capability and mechanisms responsible for the production of reactive oxygen species (ROS) by catalytic AgOx/TiOx plasma coatings fabricated room temperature have been carefully investigated. Results indicate that •OH and $O_2^{•-}$ radicals are primarily formed, followed by their recombination. The type and amount of ROS delivered can be precisely controlled at the nanoscale by functionalization with nanoporous, superhydrophilic plasma polymer layers. Varying the thickness of the np-SiOx layer from 7 to 100 nm, a consistent and specific control of the type and amount of ROS delivered can be achieved. The addition of a very thin ($d_{np-SiOx} < 10$ nm) functional layer allows the partial conversion of $O_2^{•-}$ to $^1O_2$ and subsequent delivery of both type of ROS to the outside environment. Tuning the thickness of the np-SiOx layer allows controlling short-lived to long-lived ROS recombination. ROS delivery levels are always below cytotoxicity levels, making the material suitable for applications in contact with cells such as wound healing,[7] face masks, or surgical instruments, among others. Additionally, the functional layer avoids direct contact of cells with the metal oxide surface. Results obtained for the production and delivery of ROS have been correlated with the antibacterial and antiviral properties of the catalytic material. Different bacteria and virus models have been studied, indicating that antimicrobial mechanisms are related to ROS diffusion and reactions with key macromolecules of the microorganisms. Hence, a catalytic metal oxide plasma coating combined with plasma functionalization offers a powerful toolbox to provide tailored ROS-delivering surfaces. Industrial implementation is facilitated based on established plasma technology and one-step processing.

**Methods**

*Fabrication of catalytic thin films (AgOx/TiOx):* The complete fabrication process was carried out in a pilot-scale plasma chamber equipped with a rotatable drum. The reactor contains separated radiofrequency (RF) plasma and sputtering regions, allowing for the performance of a one-step process. Silicon wafers, round glass slides, fabrics, and wound dressings were used as substrates. Substrates were cleaned with Ar plasma (160 sccm Ar at 0.1 mbar, 400 W RF-power for 10 minutes). After that, first titanium (Ti) and then silver (Ag) targets, located one in front of the other, were sputtered (in the metallic mode). Specifically, a layer of 50 nm of TiOx was sputtered from the Ti target (70 sccm Ar at 0.008 mbar, applying pulsed DC (100 kHz) - 2 kW at 320 V- for 8 minutes). Note that, although no reactive sputtering is carried out (no $O_2$ added to the plasma), nonstoichiometric TiOx is formed because of residual water and oxygen content in the chamber.[25,59] Subsequently, a nominally 5-nm-thick layer of Ag was sputtered from the Ag-target (70 sccm Ar at 0.008 mbar, applying pulsed DC (100 kHz) - 0.6 kW at 360 V- for 36 sec). As the deposited Ag is in metallic state, plasma post-oxidation is applied to oxidize to AgOx. Consequently, samples were exposed to an Ar/$O_2$ discharge in the RF-plasma zone (200 sccm Ar/200 sccm $O_2$ at 0.25 mbar, 600 W for 10 min), resulting in AgOx/TiOx with a total thickness of 55 ± 5 nm. A similar fabrication procedure has been described by Hegemann *et al.* [25].

*Functionalization of catalytic thin films with nanoporous SiOx plasma polymer films (np-SiOx/AgOx/TiOx):* AgOx/TiOx catalytic thin films were further coated with nanoporous SiOx (np-SiOx) plasma polymer films acting as functional layers to precisely control ROS delivery. Functional layers with 13 ± 2 % volumetric porosity were fabricated in a symmetric RF capacitively coupled (CCP) plasma chamber, with the samples located at the electrode and fully exposed to the plasma.[29] Cycles of plasma polymerization (Ar/$O_2$/HMDSO – depositing ~7 nm SiO:CH) and etching (Ar/$O_2$ – removing the residual hydrocarbon content over 5 min) were carried out, varying the coating thickness between 7 and 100 nm. Similar layers with a higher volumetric porosity (20 ± 5 %) were also fabricated following a similar procedure in the sample plasma chamber used for the fabrication of the AgOx/TiOx coatings. In this case, the samples were located at the wall of the reactor and coated following a method to reduce ion bombardment (near-plasma chemistry approach).[30]

*Characterization of AgOx/TiOx and functionalizing thin films*: P-doped silicon wafers were used as substrates for profilometry and morphology characterization. The thickness of the deposited layers was measured using a Bruker Dektak XT profilometer. SEM micrographs were acquired in top view with a Hitachi S4800 microscope operated at 2 kV detecting secondary electrons. AFM analysis was conducted using a Bruker Dimension Icon AFM (Billerica, US). The AFM was operated in tapping mode with an Olympus OMCL-AC160TS-R3 probe, characterized by a resonant frequency of 300 kHz, a spring constant of 26 N/m, and a nominal tip radius of 7 nm. The obtained images were processed and analyzed using NanoScopeAnalysis software (version 3.00). A polynomial flatten filter (n=1) and a plane fit (polynomial n=1, XY mode) were applied to all presented AFM images. ATR-FTIR (Varian 640-IR, Agilent) was used to characterize the chemistry of the functional layers (i.e., plasma polymer films), with the samples deposited on aluminum foil. Optical characterization of the polymers was carried out in a UV-Vis Cary 4000 (Agilent) spectrometer. Ellipsometry was used to determine the refractive index and, therefore, the porosity of PPFs, using a Nanofilm EP4 device (Accurion); the device was operated at a constant wavelength of 658 nm and varying the angle of incidence between 55 ° and 80 ° (1° step). For a more detailed explanation see our previous publications.[29,30]

*ROS detection by fluorescence*: The production of reactive oxygen species (ROS) was monitored in a Biotek Synergy H1 Multimode Reader (Agilent), using specific fluorescence excitation and emission wavelengths. The instrument was operated at normal speed, with a delay of 100 ms and measuring from top. The measurements were conducted by introducing the samples deposited on glass (10 mm diameter) in a 48 well-plate and pipetting 300 µL of water-based solutions of the specific dyes. The reproducibility of the results has been checked also working with a 96-well plate, 6 mm samples and pipetting 100 µL solutions. The signal of the glass substrate, as well as the dye itself, were controlled during the measurements. The following molecules have been used for the detection: dihydrorhodamin 123 (DHR 123, Sigma Aldrich), at concentration 10 µM, following its oxidation into rhodamine 123 with excitation at 488 nm and emission at 535 nm; dihydroethidium (DHE, Sigma Aldrich), at concentration 10 µM, following the degradation of the dye itself with excitation at 355 nm and emission at 420 nm; 9,10-Anthracenediyl-bis(methylene)dimalonic acid (ABDA), at concentration 2.4 µM, with excitation at 404 nm and emission at 490 nm. To ensure reproducibility, experiments were conducted more than three times, with a minimum of three replicates per experiment.

*ROS detection by spin trapping EPR:* Catalytic plasma coatings were incubated in a 24-well plate with 200 µL of 100 mM DMPO (Dojindo) in 20 mM HEPES, 150 mM NaCl buffer at pH 7.3, under continuous light irradiation with an LED lamp (3.4 W, 220 V). All the kinetics of DMPO·-OH adduct formation were performed at room temperature. Approximately every 10 minutes (with an offset of 3 minutes for the first point), 20 µL of solution were collected in a glass tube of 0.7 mm i.d. and measured with a Bruker E500 X-band cw EPR spectrometer, coupled with an ER 4122 SHQ cavity. For each measurement, 80 G were swept in 60 s with a time constant of 40.96 ms at a microwave power of 9.85 mW. The modulation was set to 1.5 G of amplitude and 100 kHz of frequency. For each measurement, 4 scans were averaged, and then the sample was recollected from the glass tube and pipetted back in the well plate.

*Cytotoxicity:* Possible cytotoxicity of the catalytic plasma coatings was analyzed using normal human dermal fibroblasts (NHDFs; C-12352; PromoCell). Samples were extracted in 150 µL DMEM (Dulbecco's Modified Eagle Medium) containing 1% L-Glutamine, 1% penicillin/streptomycin, and 10% fetal calf serum (FCS). Empty wells without any sample were used as negative controls. The extraction process was carried out at 37°C with 100% humidity and 5% $CO_2$ for 24h. NHDFs were seeded with 10 000 cells per well (TPP Techno Plastic Products AG, Trasadingen, Switzerland) in 100 µL DMEM supplemented with 1% L-Glutamine, 1% penicillin/streptomycin and 10% fetal calf serum 1 day before incubation with extracts. The NHDFs were then incubated for 24 h with 100 µL extracts by replacing the old media. The viable NHDFs cells of negative control were set as 100%, and the ones incubated with 1% Triton X-100 in supplemented DMEM were regarded as the positive control. Cell viability of the NHDFs was determined via MTS assay by measuring the absorbance at 490 nm.

*Skin irritation and sensitization:* A KeratinoSens® skin sensitization & irritation assay was performed according to OECD guideline No. 442D [60], described in detail described in [61], in order to assess the acute in vitro skin toxicity of the catalytic plasma coatings (np-SiOx/AgOx/TiOx; AgOx/TiOx) substrates. To mimic a more realistic exposure scenario, artificial sweat [61] was applied as an extract solvent during an extraction conducted at 37°C for 24h (± 1h) on an orbital shaker set to 300rpm in darkness. Extraction volume ratio is determined according to ISO10993-12.[62] In short, positive control: EGDMA (1 – 1'000 µM) and artificial sweat catalytic plasma coatings sample extracts (0.005 – 10%) were contacted for 48h onto KeratinoSens® skin cells maintained in DMEM assay media in 96 well plates. Skin irritation (cell viability) was assessed with AlamarBlue (Invitrogen, DAL1100, excitation 540nm / emission 590nm) whereas skin sensitization (fold-induction) was evaluated with One-Glo™ Reagent (Promega, E6120, integration time = 1 s/well). Subsequently, $EC_{1.5}$ values (fold-induction) and $IC_{50}$ values of dose-dependent controls and samples were calculated according to OECD guideline No. 442D.[60]

*Antibacterial activity*: The antibacterial activity of the catalytic plasma coatings was qualitatively assessed by touch test and quantitatively by a modified ASTM method. The samples were deposited on glass (10 mm). *Escherichia coli* (E. coli) DSM 1003 and *Staphylococcus aureus* (S. aureus) ATCC 6538 were selected as gram-positive and gram-negative representative pathogens, respectively. Three replicates were analyzed for each measurement. For the qualitative (touch test) antibacterial assay, bacteria colonies from the respective agar plate stock were cultured overnight in 5 mL Tryptic Soy Broth (TSB) supplemented with 0.25% Glucose media at 37 °C with agitation (160 rpm). The overnight culture was diluted with TSB to 0.1 of optical density at 600 nm (OD600) and regrown for 1.5-2 hours to reach exponential growth. Thereafter, the regrown cultures were diluted in TSB to a final OD600nm of 0.05 for *S. aureus* and of 0.1 for *E. coli*. 100 µL of the resulting suspension was plated on Plate Count Agar (PC-agar) and let dry for 30 min in the air. The catalytic plasma coating samples were placed onto these agar plates, and allowed interaction for 10 and 60 min. The agar plates were then incubated at 37°C for overnight, and the resultant growth inhibition was visualized and documented photographically using SCAN300 (Interscience, France).

The antibacterial efficacy was further measured quantitatively for both bacterial strains following a modified ASTM method.[25,47] Bacteria colonies from the respective agar plate stock were again cultured overnight in 5 mL TSB and glucose media as described above. After overnight culture and reaching exponential growth, regrown cultures were diluted to a final OD600nm of 0.01 in 0.9% NaCl. Therefrom, 100 µL were loaded onto the samples as well as the controls. The samples were then incubated for 60 minutes at room temperature without shaking. The suspension was subsequently removed, and the samples were washed twice with 1 mL 0.9% NaCl to remove the adhered bacteria. The removed bacterial suspension and the washing solution from each sample was collected and mixed. The washed samples were placed in 2.5 ml 1x PBS, sonicated for 5 min, and thereafter vigorously vortexed for 15 s. Serial dilutions of the two collected bacterial mixtures were spotted on PC agar plates, then incubated at 37°C overnight. After incubation, bacterial colonies were counted to obtain an estimation of the viable cells on each of the samples.

*Antiviral InVIS membrane integrity:* The inactivated virus membrane integrity assay was performed according to the literature [50] with minor modifications. AgOx/TiOx and np-SiOx/AgOx/TiOx samples deposited on glass (1 cm diameter) were tested, as well as bare glass substrates and the addition of positive control (OEG, 1.25 mg mL$^{-1}$). In short, 20 µL InVIS solution was dropcasted onto the samples and incubated for 20 min. Then, each sample was transferred into falcon tubes containing 20 mL of PBS and underwent agitation to ensure the homogenization of the freshly added PBS and the remaining inoculum. After mixing, 2 mL of the washout was collected and pipetted into transparent cuvettes. The emission fluorescence spectra of the washout were characterized using the Horiba FluoroMax SpectraFluorometer. Excitation wavelength was set to 560 nm and the emission intensity was measured between 580 and 650 nm.

*Antiviral MHV activity:* The antiviral activity was evaluated according to ISO 18184:2019(E) [63], described in more detail in [61]. This assessment focused on the inactivation potential of the catalytic plasma coatings (AgOx/TiOx and np-SiOx/AgOx/TiOx) against the murine hepatitis virus (Coronavirus, Group 2, ATCC® VR1426™) over an exposure period of 2 hours. Plaque assays applying L929 murine subcutaneous connective tissue host cells were conducted in independent triplicates (N=3, n=2). Negative controls and immediate washout were performed with non-plasma treated glass surface carriers. Subsequently, in accordance with ISO 18184 guideline,[63] the antiviral activity was determined.

**Supporting Information**

Supporting Information includes: S1 (FIB-cross section SEM analysis of np-SiOx/AgOx/TiOx), S2 (Total/diffuse transmittance and diffuse reflectance data), S3 (Pseudo-first order degradation kinetic analysis of DHE), S4 (DHE degradation experiment in anoxic conditions), S5 (Inhibition experiment incorporating GSH in DHE solution), S6 and S7 (Additional EPR results).

**Acknowledgements**

P.N., Q.R., and D.H. acknowledge funding by SNSF (COST 2022, project 213368). M.G. acknowledges funding by SNSF (project grant No. 200021_204111 (Np Surf)). G.R. and P.M. thank EU-NOVA project, the European Union's Horizon Europe Framework Programme under grant agreement number 101058554. This work was co-funded by the Swiss State Secretariat for Education, Research and Innovation (SERI) and the UK Research and Innovation (UKRI) under the UK government's Horizon Europe funding guarantee grant number 10042534 & grant number 10055606.

The authors acknowledge Josua Roduner, Elin Fehr, Emmanuel Chabal, and Julia Morini for their assistance, and Christian Zaubitzer from the ScopeM of the ETH Zürich for the facilities. P.N. and F.K. thank Prof. Thomas Bürgi (University of Geneva) for facilitating Kalemi's internship at Empa. P.N. acknowledges Carmen Gutierrez Lázaro and Elena Cabello Olmo (Materials Science Institute of Seville) for their assistance and advice for materials optical characterization, and Ana Borrás and Carmen López Santos from the same institution for the fruitful discussions and their support.

**Conflict of Interest**

The authors declare no conflict of interest.

**Data Availability Statement**

Data shown in this study are available from the corresponding authors upon reasonable request.

**References**

[1] Y. Chong, Q. Liu, C. Ge, *Nano Today* **2021**, *37*, 101076.
[2] A. Gomes, E. Fernandes, J. L. F. C. Lima, *Journal of Biochemical and Biophysical Methods* **2005**, *65*, 45.
[3] M. Dryden, *International Journal of Antimicrobial Agents* **2018**, *51*, 299.
[4] M. Huo, L. Wang, Y. Chen, J. Shi, *Nat Commun* **2017**, *8*, 357.
[5] M. B. Chobba, M. L. Weththimuni, M. Messaoud, C. Urzi, J. Bouaziz, F. De Leo, M. Licchelli, *Progress in Organic Coatings* **2021**, *158*, 106342.
[6] R. Ghosh, A. Baut, G. Belleri, M. Kappl, H.-J. Butt, T. M. Schutzius, *Nat Sustain* **2023**, 1.
[7] C. Dunnill, T. Patton, J. Brennan, J. Barrett, M. Dryden, J. Cooke, D. Leaper, N. T. Georgopoulos, *International Wound Journal* **2017**, *14*, 89.
[8] Z. Zhou, B. Li, X. Liu, Z. Li, S. Zhu, Y. Liang, Z. Cui, S. Wu, *ACS Appl. Bio Mater.* **2021**, *4*, 3909.
[9] M. Živanić, A. Espona-Noguera, H. Verswyvel, E. Smits, A. Bogaerts, A. Lin, C. Canal, *Advanced Functional Materials* **2024**, *34*, 2312005.
[10] X. Dong, W. Wu, P. Pan, X.-Z. Zhang, *Advanced Materials* **2023**, 2304963.
[11] L. R. H. Gerken, A. Gogos, F. H. L. Starsich, H. David, M. E. Gerdes, H. Schiefer, S. Psoroulas, D. Meer, L. Plasswilm, D. C. Weber, I. K. Herrmann, *Nat Commun* **2022**, *13*, 3248.
[12] C. Zhang, X. Wang, J. Du, Z. Gu, Y. Zhao, *Advanced Science* **2021**, *8*, 2002797.
[13] V. Rico, P. Romero, J. L. Hueso, J. P. Espinós, A. R. González-Elipe, *Catalysis Today* **2009**, *143*, 347.
[14] B. C. Hodges, E. L. Cates, J.-H. Kim, *Nature Nanotech* **2018**, *13*, 642.
[15] J. Zambrano, R. Irusta-Mata, J. J. Jiménez, S. Bolado, P. A. García-Encina, in *Development in Wastewater Treatment Research and Processes* (Eds.: M. Shah, S. Rodriguez-Couto, J. Biswas), Elsevier, **2022**, pp. 543–572.

[16] C. Cao, X. Wang, N. Yang, X. Song, X. Dong, *Chem. Sci.* **2022**, *13*, 863.

[17] U. Chilakamarthi, L. Giribabu, *The Chemical Record* **2017**, *17*, 775.

[18] X. Liu, Y. Yang, M. Ling, R. Sun, M. Zhu, J. Chen, M. Yu, Z. Peng, Z. Yu, X. Liu, *Advanced Functional Materials* **2021**, *31*, 2101709.

[19] L. Jacquemin, Z. Song, N. Le Breton, Y. Nishina, S. Choua, G. Reina, A. Bianco, *Small* **2023**, *19*, 2207229.

[20] I. Adamovich, S. Agarwal, E. Ahedo, L. L. Alves, S. Baalrud, N. Babaeva, A. Bogaerts, A. Bourdon, P. J. Bruggeman, C. Canal, E. H. Choi, S. Coulombe, Z. Donkó, D. B. Graves, S. Hamaguchi, D. Hegemann, M. Hori, H.-H. Kim, G. M. W. Kroesen, M. J. Kushner, A. Laricchiuta, X. Li, T. E. Magin, S. M. Thagard, V. Miller, A. B. Murphy, G. S. Oehrlein, N. Puac, R. M. Sankaran, S. Samukawa, M. Shiratani, M. Šimek, N. Tarasenko, K. Terashima, E. T. Jr, J. Trieschmann, S. Tsikata, M. M. Turner, I. J. van der Walt, M. C. M. van de Sanden, T. von Woedtke, *J. Phys. D: Appl. Phys.* **2022**, *55*, 373001.

[21] K. J. Heo, S. B. Jeong, J. Shin, G. B. Hwang, H. S. Ko, Y. Kim, D. Y. Choi, J. H. Jung, *Nano Lett.* **2021**, *21*, 1576.

[22] E. Bletsa, P. Merkl, T. Thersleff, S. Normark, B. Henriques-Normark, G. A. Sotiriou, *Chemical Engineering Journal* **2023**, *454*, 139971.

[23] J. Liu, L. Ye, S. Wooh, M. Kappl, W. Steffen, H.-J. Butt, *ACS Appl. Mater. Interfaces* **2019**, *11*, 27422.

[24] X. Pan, M.-Q. Yang, X. Fu, N. Zhang, Y.-J. Xu, *Nanoscale* **2013**, *5*, 3601.

[25] D. Hegemann, B. Hanselmann, F. Zuber, F. Pan, S. Gaiser, P. Rupper, K. Maniura-Weber, K. Ruffieux, Q. Ren, *Plasma Processes & Polymers* **2022**, *19*, 2100246.

[26] S. A. Abdullah, M. Z. Sahdan, N. Nafarizal, H. Saim, Z. Embong, C. H. Cik Rohaida, F. Adriyanto, *Applied Surface Science* **2018**, *462*, 575.

[27] H. Cao, Y. Qiao, X. Liu, T. Lu, T. Cui, F. Meng, P. K. Chu, *Acta Biomaterialia* **2013**, *9*, 5100.

[28] P. Navascués, M. Buchtelová, L. Zajíčková, P. Rupper, D. Hegemann, *Applied Surface Science* **2024**, *645*, 158824.

[29] T. Gergs, C. Monti, S. Gaiser, M. Amberg, U. Schütz, T. Mussenbrock, J. Trieschmann, M. Heuberger, D. Hegemann, *Plasma Processes and Polymers* **2022**, *19*, 2200049.

[30] P. Navascués, U. Schütz, B. Hanselmann, D. Hegemann, *Nanomaterials* **2024**, *14*, 195.

[31] A. J. Haider, Z. N. Jameel, I. H. M. Al-Hussaini, *Energy Procedia* **2019**, *157*, 17.

[32] S. Rtimi, S. Giannakis, R. Sanjines, C. Pulgarin, M. Bensimon, J. Kiwi, *Applied Catalysis B: Environmental* **2016**, *182*, 277.

[33] H. Kumar Raut, V. Anand Ganesh, A. Sreekumaran Nair, S. Ramakrishna, *Energy & Environmental Science* **2011**, *4*, 3779.

[34] T. Liu, L. Sun, Y. Zhang, Y. Wang, J. Zheng, *Journal of Biochemical and Molecular Toxicology* **2022**, *36*, e22942.

[35] X. Ding, K. Zhao, L. Zhang, *Environ. Sci. Technol.* **2014**, *48*, 5823.

[36] T. Entradas, S. Waldron, M. Volk, *Journal of Photochemistry and Photobiology B: Biology* **2020**, *204*, 111787.

[37] F. A. Villamena, *Reactive Species Detection in Biology. From Fluorescence to Electron Paramagnetic Resonance Spectroscopy*, Elsevier, **2016**.

[38] Y. Wang, Y. Lin, S. He, S. Wu, C. Yang, *Journal of Hazardous Materials* **2024**, *461*, 132538.

[39] A. V. Demyanenko, A. S. Bogomolov, N. V. Dozmorov, A. I. Svyatova, A. P. Pyryaeva, V. G. Goldort, S. A. Kochubei, A. V. Baklanov, *J. Phys. Chem. C* **2019**, *123*, 2175.

[40] C. M. C. Andrés, J. M. Pérez de la Lastra, C. Andrés Juan, F. J. Plou, E. Pérez-Lebeña, *Int J Mol Sci* **2023**, *24*, 1841.

[41] R. Ueki, Y. Imaizumi, Y. Iwamoto, H. Sakugawa, K. Takeda, *Science of The Total Environment* **2020**, *716*, 136971.

[42] K. Krumova, G. Cosa, in *Singlet Oxygen: Applications in Biosciences and Nanosciences*, **2016**.

[43] M. R. Hoffmann, S. T. Martin, W. Choi, D. W. Bahnemann, *Chem. Rev.* **1995**, *95*, 69.

[44] Y. Chen, C. Shen, J. Wang, G. Xiao, G. Luo, *ACS Sustainable Chem. Eng.* **2018**, *6*, 13276.

[45] J. Du, C. Wang, Z. Zhao, F. Cui, Q. Ou, J. Liu, *Chemical Engineering Science* **2021**, *241*, 116683.

[46] E. Bülbül, D. Hegemann, T. Geue, M. Heuberger, *Colloids and Surfaces B: Biointerfaces* **2020**, *190*, 110908.

[47] F. Pan, S. Altenried, F. Zuber, R. S. Wagner, Y.-H. Su, M. Rottmar, K. Maniura-Weber, Q. Ren, *Colloids and Surfaces B: Biointerfaces* **2021**, *206*, 111940.

[48] J. Bogdan, J. Zarzyńska, J. Pławińska-Czarnak, *Nanoscale Res Lett* **2015**, *10*, 309.

[49] T. B. Cavitt, N. Pathak, *Pharmaceuticals* **2021**, *14*, 977.

[50] L. A. Furer, P. Clement, G. Herwig, R. M. Rossi, F. Bhoelan, M. Amacker, T. Stegmann, T. Buerki-Thurnherr, P. Wick, *Sci Rep* **2022**, *12*, 11583.

[51] S. L. Warnes, E. N. Summersgill, C. W. Keevil, *Applied and Environmental Microbiology* **2015**, *81*, 1085.

[52] S. Ye, K. Shao, Z. Li, N. Guo, Y. Zuo, Q. Li, Z. Lu, L. Chen, Q. He, H. Han, *ACS Appl. Mater. Interfaces* **2015**, *7*, 21571.

[53] P. D. Rakowska, M. Tiddia, N. Faruqui, C. Bankier, Y. Pei, A. J. Pollard, J. Zhang, I. S. Gilmore, *Commun Mater* **2021**, *2*, 1.

[54] N. Kaushik, S. Mitra, E. J. Baek, L. N. Nguyen, P. Bhartiya, J. H. Kim, E. H. Choi, N. K. Kaushik, *Journal of Advanced Research* **2023**, *43*, 59.

[55] Y. Chen, X. Tang, X. Gao, B. Zhang, Y. Luo, X. Yao, *Ceramics International* **2019**, *45*, 15505.

[56] S. Jung, D.-H. Kim, H.-S. Ahn, H.-J. Go, Z. Wang, D. Yeo, S. Woo, Y. Seo, M. I. Hossain, I.-S. Choi, S.-D. Ha, C. Choi, *Food Control* **2023**, *143*, 109306.

[57] M. Z. Waleed, K. Rafiq, M. Z. Abid, M. Burhan, R. H. Althomali, S. Iqbal, E. Hussain, *Journal of Environmental Chemical Engineering* **2024**, *12*, 112713.

[58] K. Rule Wigginton, L. Menin, J. P. Montoya, T. Kohn, *Environ. Sci. Technol.* **2010**, *44*, 5437.

[59] M. Amberg, P. Rupper, R. Storchenegger, M. Weder, D. Hegemann, *Nanomedicine: Nanotechnology, Biology and Medicine* **2015**, *11*, 845.

[60] In *OECD/OCDE. Test Guideline No. 442D. Test No. 442D: In Vitro Skin Sensitisation: ARE-Nrf2 Luciferase Test Method, OECD Guidelines for the Testing of Chemicals. Volume 4*, **2022**.

[61] P. Meier, P. Clement, S. Altenried, G. Reina, Q. Ren, R. Züst, O. Enger, F. Choi, N. Nestle, T. Deisenroth, P. Neubauer, P. Wick, *Sci Rep* **2023**, *13*, 20556.

[62] *ISO Standard 10993-12. Biological Evaluation of Medical Devices—Part 12: Sample Preparation and Reference Materials. In ISO 10993-12. Volume 4*, **2012**.

[63] *ISO Standard 18184:2019. Textiles—Determination of Antiviral Activity of Textile Products. In ISO 18184:2019(E). Volume 6*, **2019**.

Table of contents (TOC)

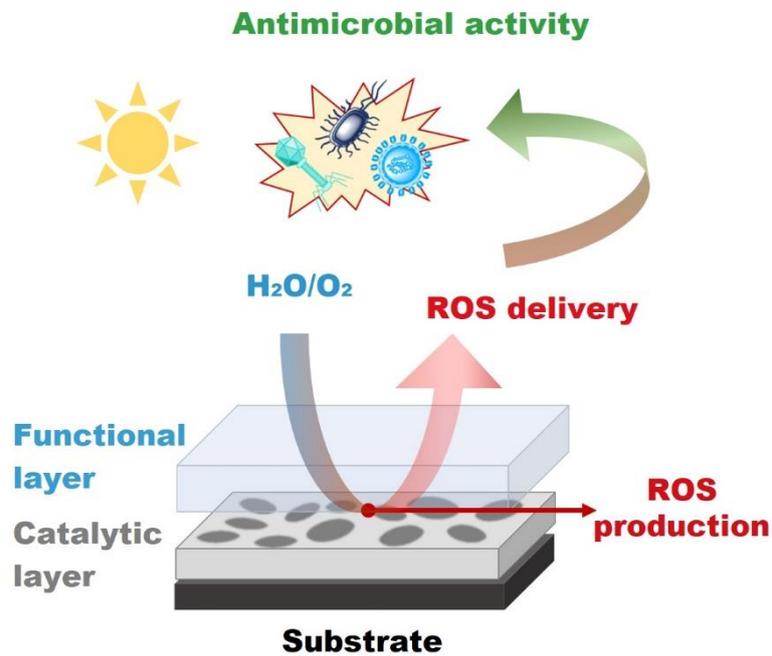

The amount and type of ROS delivered by photocatalytic surfaces are controlled by nanoscale surface functionalization. Based on the reactive species' lifetime, superoxide anion and singlet oxygen release are adjusted by tuning the functional layer thickness. All layers in the thin film material are fabricated by low-pressure plasma technology. The ROS-releasing surfaces show antimicrobial activity, avoiding cytotoxic and sensitization effects.